\documentclass[useAMS,usenatbib]{mn2e}
\usepackage{times}
\usepackage{amssymb}
\usepackage{graphicx}
\usepackage{rotating}
\usepackage{lscape}
\usepackage{multirow}

\begin{document}

\title[Near-infrared evolution of BCGs since z=1]{Near-infrared evolution of brightest
cluster galaxies in the most X-ray luminous clusters since z=1}
\author[J.P.Stott et al.]{J. P. Stott$^{1, 2}$\thanks{E-mail: jps@astro.livjm.ac.uk}, A. C. Edge$^1$, G. P. Smith$^{3, 4}$, A. M. Swinbank$^1$, H. Ebeling$^5$\\
\\
$^1$Institute for Computational Cosmology, Department of Physics, University of Durham, South Road, Durham DH1 3LE, UK\\
$^2$Astrophysics Research Institute, Liverpool John Moores University, Twelve Quays House, Egerton Wharf, Birkenhead, CH41 1LD, UK\\
$^3$California Institute of Technology, Mail Code 105-24,
  Pasadena, CA 91125, USA.\\ 
$^4$  School of Physics $\&$ Astronomy, University of Birmingham, Edgbaston, Birmingham, B15 2TT, UK \\
$^5$ Institute for Astronomy, 2680 Woodlawn Drive, Honolulu, HI 96822, USA\\}

\date{}

\pagerange{\pageref{firstpage}--\pageref{lastpage}} \pubyear{2006}

\maketitle

\label{firstpage}

\begin{abstract}
We investigate the near infrared evolution of brightest cluster galaxies (BCGs) from
a sample of rich galaxy clusters since $z\sim1$. By employing an X-ray selection of $L_{X}>10^{44} $erg s$^{-1}$ we limit environmental effects by selecting BCGs in comparably high density regions. We find a positive relationship between X-ray and near-infrared luminosity for BCGs in clusters with $L_{X}>5\times10^{44} $erg s$^{-1}$. Applying a correction for this relation we reduce the scatter in the BCG absolute magnitude by a factor of 30\%. The near-infrared $J-K$ colour evolution demonstrates that the stellar population in BCGs has been in place since at least $z=$2 and that we expect a shorter period of star formation than that predicted by current hierarchical merger models. We also confirm that there is a relationship between `blue' $J-K$ colour and the presence of BCG emission lines associated with star formation in cooling flows.
\end{abstract}


\section{Introduction}
\label{sec:intro}
A Brightest Cluster Galaxy (BCG) is a giant elliptical galaxy near the spatial and gravitational centre of a galaxy cluster. BCGs are the brightest and most massive stellar systems in the Universe. Their high luminosities and small scatter in absolute magnitude makes them effective standard candles. As such they were originally used by astronomers to confirm and considerably  increase the range of Hubble's redshift - distance law (e.g ~\citealt{Sandage1972}). BCGs are particularly important for galaxy formation and evolution studies as the above properties make them less prone to selection effects and biasing. Near-infrared photometry is often chosen for BCG studies as K correction, stellar evolution and extinction by dust in this region of the spectrum are considerably less than at optical wavelengths. 

There is considerable observational evidence that suggests giant ellipticals were formed at high redshift and have been passively evolving to the present day \citep{Bower1992,AragonSal1993,vanDok1998,Stanford1998}. Passive evolution describes a situation where the stellar population in a galaxy forms in a single burst at a redshift $z_{f}$. This population then matures, without further star formation. No evolution describes the case where the observed luminosity changes over cosmic time of a stellar population are purely attributed to the effects of distance and K correction. Depending on the cluster selection technique the BCG photometry can follow drastically different evolutionary tracks. For example highly luminous X-ray clusters tend to prefer evolving models whereas low L$_{X}$ clusters are seen to have stellar populations preferring no evolution (\citealt{Alfonso1998}, ~\citealt{BCM2000} and ~\citealt{Nelson2002}). 

The latest hierarchical simulations of BCG formation predict that the stellar components of BCGs are formed very early (50 per cent at $z\sim$5 and 80 per cent at $z\sim$3, \citealt{DeLucia2007}). This star formation occurs in separate sub-components which then accrete to form the BCG through `dry' mergers. It is important to note that in these simulations local BCGs are not directly descended from high-$z$ ($z>0.7$) BCGs. However, \cite{DeLucia2007} find little physical difference between the progenitors of local BCGs and high-$z$ BCGs or between the local BCGs and the descendants of the high-$z$ BCGs. This means that observed evolution presented here can still be compared to simulation.

In this paper we aim to test the above results and provide further constraints to simulations by comparing the K band and $J-K$ colour evolution of a well defined X-ray selected sample of BCGs to a set of evolution models. 

We study a large sample of the most X-ray luminous clusters known which correspond to the most extreme environments at their respective epochs. The motivation for studying an X-ray selected sample of clusters is to ensure that we are observing objects in similar high mass, high density environments. This homogeneity is key to our study as we wish to compare clusters over a range of redshifts. By incorporating clusters from the MAssive Cluster Survey (MACS, ~\citealt{MACS2001}) we are going to higher X-ray luminosity than any previous BCG study.

Lambda CDM cosmology ($\Omega_{M}=$0.3, $\Omega_{Vac}=$0.7, $H_{0}=$70) and the Vega magnitude system are used throughout.

\section{Data}
\label{sec:data}
\subsection{The sample}

To select BCGs in a homogeneous sample of massive clusters from $z=$0 -- 1 we require X-ray selected clusters from a number of large surveys. These clusters are all selected to have X-ray luminosities in excess of $10^{44} $erg s$^{-1}$ (0.1 -- 2.4 keV) and therefore correspond to the most massive clusters known. The $z<$0.3 sample are taken from the \emph{ROentgen SATellite} (\emph{ROSAT}) Brightest Cluster Survey (BCS), extended BCS ~\citep{Ebeling2000} and the X-ray Brightest Abell Clusters Survey (XBACS, \citealt{xbacs1996}). We then select a comparable sample at $0.3<z<0.7$ from the MAssive Cluster Survey (MACS, ~\citealt{MACS2001}). Additional high redshift clusters are sourced from analysis of archival observations of the clusters MS1054-0321 and RCS0224-0002. Details of the sample can be found in tables \ref{tab:sample2m}, \ref{tab:samplew} and \ref{tab:sampleu}. 

Our sample also contains 4 additional BCGs from spectroscopically confirmed $z\sim$0.9 optical-infrared selected clusters discovered in \citealt{Swin2007}. These BCGs where found in the UKIRT Infrared Deep Sky Survey (UKIDSS, \citealt{Law2007}) Deep eXtragalactic Survey (DXS, Survey Head: Alastair Edge). They are not yet confirmed as high $L_{X}$ clusters but they do have absolute magnitudes comparable with the rest of our sample (mean absolute magnitude of DXS BCGs is -26.6). 

\begin{figure}
\centering
\scalebox{0.50}[0.50]{\includegraphics*{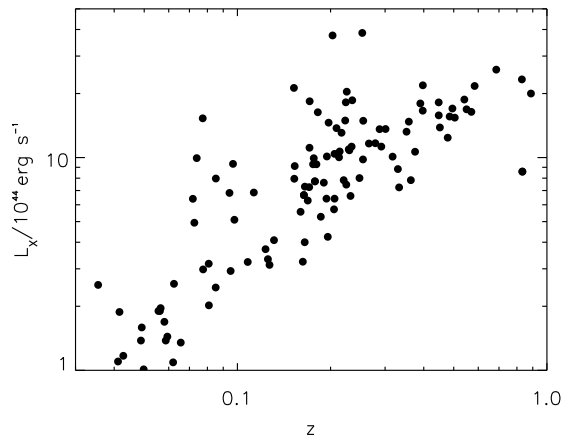}}
\caption[]{The X-ray luminosity vs $z$ for our sample. }
\label{fig:xz}
\end{figure}

Fig. \ref{fig:xz} shows the X-ray luminosity vs redshift for our sample which demonstrates that we are going to higher X-ray luminosity than any previous BCG study as we have excellent coverage in the $L_{X}>10^{45} $erg s$^{-1}$ range. In total we have a sample of 121 BCGs available for analysis of which 47 are in the $L_{X}>10^{45} $erg s$^{-1}$ regime compared to only 7 from the \cite{BCM2000} sample.

\subsection{Photometry}
To study a sample of BCGs spanning such a wide redshift range we obtain data from several sources. 

A combination of the Two Micron All Sky Survey (2MASS) extended and point source catalogues ~\citep{2mass2006} is used for the $z<$0.15 BCS, eBCS and XBACS BCGs. The limiting magnitudes for the XSC are $J=$15.1, and $K_{s}=$13.5 mag. The eXtended Source Catalogue (XSC) K20 fiducial elliptical total magnitudes are used throughout incorporating the same aperture size in both $J$ and $K$ bands which is crucial to ensure precise colour photometry.  

The 0.15$<z<$0.3 BCS, eBCS and XBACS observations were performed in 2004 and 2005 in variable seeing (0.9" -- 1.5") with the Wide field
InfraRed Camera (WIRC, \citealt{wilson2003}) instrument on the Palomar 200" Hale telescope. These data were reduced with the WIRCTASK IRAF scripts.

The 0.3$<z<$0.7 MACS data are from 3 sources. Part of these data were obtained in 2002 in 0.4" -- 0.7" seeing using the UKIRT Fast-Track Imager (UFTI) camera on the United Kingdom InfraRed Telescope (UKIRT). The data were reduced using the ORAC-DR pipeline. More MACS clusters were observed in $\sim$1.0" seeing in 2002 again with the WIRC instrument on the Palomar 200" Hale telescope. This was reduced with the WIRCTASK IRAF scripts. The remaining clusters were observed in 2004 (P.I: J.-P. Kneib) in $\sim$0.6" seeing conditions using the Infrared Spectrometer And Array Camera (ISAAC) on the Very Large Telescope (VLT). These data were reduced with the ISAAC eclipse pipeline.

The high redshift ($z>$0.7) clusters MS1054-0321 and RCS0224-0002 are sourced from archival data. The MS1054-0321 data are from VLT/ISAAC observations obtained as part of the Faint InfraRed Extragalactic Survey (FIRES, \citealt{fires2006}) and the RCS0224-0002 cluster was observed using WIRC/Palomar.

We include additional high redshift photometry from the literature. The near-infrared selected $z\sim0.9$ clusters are sourced from UKIDSS DXS data described in \cite{Swin2007} and reanalysed here to ensure homogeneity. 

The BCG photometry for all of our z$>$0.15 near-infrared data is extracted using SExtractor's `Best' magnitude ~\citep{sextractor1996}, as used by \cite{Nelson2002}, which is comparable to the 2MASS K20 fiducial elliptical magnitude \citep{Elston2005}. The centre of clusters can be very densely populated so for crowded objects the `Best' magnitude uses the isophotal magnitude which excludes the light from close neighbours and is therefore more reliable than a fixed aperture. The `Best' magnitude is shown to be robust to galaxy shape as we find no trend between ellipticity of the aperture/BCG and the absolute K band magnitude. It is important to note that the `Best' magnitude may underestimate the integrated brightness by up to a tenth of a magnitude for BCGs at K=17.5 \citep{Martini2001} which will be the dominant error in our high redshift photometry. We run SExtractor in dual mode with the K-band apertures used to extract the J-band photometry to ensure good colour determination. The photometry calibration for our data was achieved with a combination of 2MASS and/or standard star observations.

The Wide Angle \emph{ROSAT} Pointed Surveys (WARPS, \citealt{WARPS1997}, \citealt{WARPS1998}) X-ray selected $z\sim0.9$ `total' magnitude photometry was sourced from data described in \cite{Ellis2004}.

All magnitudes are corrected for Galactic extinction using ~\cite{Schlegel1998}. The typical extinction values for our clusters were in the range 0.01 -- 0.04 mag in $K$.

\section{Analysis and Results}
\subsection{The BCG Degree of Dominance}
\label{sec:dom}
\begin{figure}
\centering
\scalebox{0.50}[0.50]{\includegraphics*{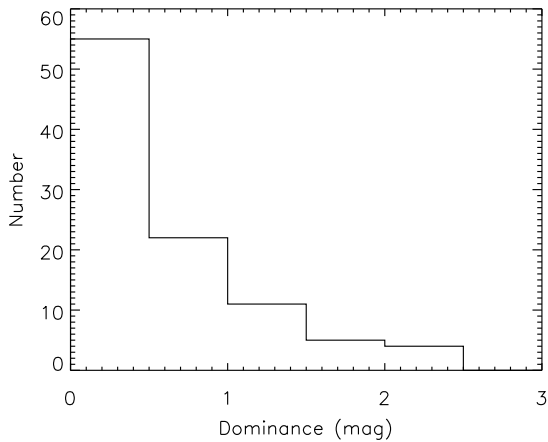}}
\caption[]{A histogram of the degree of BCG dominance for our sample. }
\label{fig:dom}
\end{figure}

We first look to see if differing cluster core environments effect our results. A measure of environment which can easily be extracted from photometric data is the degree of dominance. This parameterises the difference in luminosity between the BCG and the next brightest galaxies in the cluster. The BCG may be the dominant elliptical in a cluster centre containing much smaller galaxies or it may be in a system were it is only marginally brighter than the next brightest members. The degree of dominance is defined as $\Delta m_{1-2,3}= (m_2+m_3)/2 -m_1$ where $m_1$ is the magnitude of the BCG and $m_2$ and $m_3$ are the magnitudes of the 2nd and 3rd brightest members respectively ~\citep{Kim2002}. The 2nd and 3rd brightest galaxies are selected as the next two brightest galaxies on the cluster red sequence within a radius of 500kpc of the BCG. Taking the average of the 2nd and 3rd ranked galaxies is slightly more robust to contamination than just using the 2nd. It also removes the weighting from cases where there are two BCG candidates that are far more luminous than the rest of the cluster. Fig. \ref{fig:dom} shows a histogram of the distribution of the degree of $K$ band dominance for our sample. The maximum cluster dominance found in our sample is 2.43 mag. We find that the majority of our BCGs are in cluster environments where they are not highly dominant. The key result is that we find no trend between dominance and redshift, X-ray luminosity or absolute K band magnitude for our sample. We are therefore satisfied that the differing galaxy environment between cluster cores has no effect on the results presented in this paper. 

\subsection{$K$ Band Correction}
\label{sec:xc}
\begin{figure*}
\centering
\scalebox{0.50}[0.50]{\includegraphics*{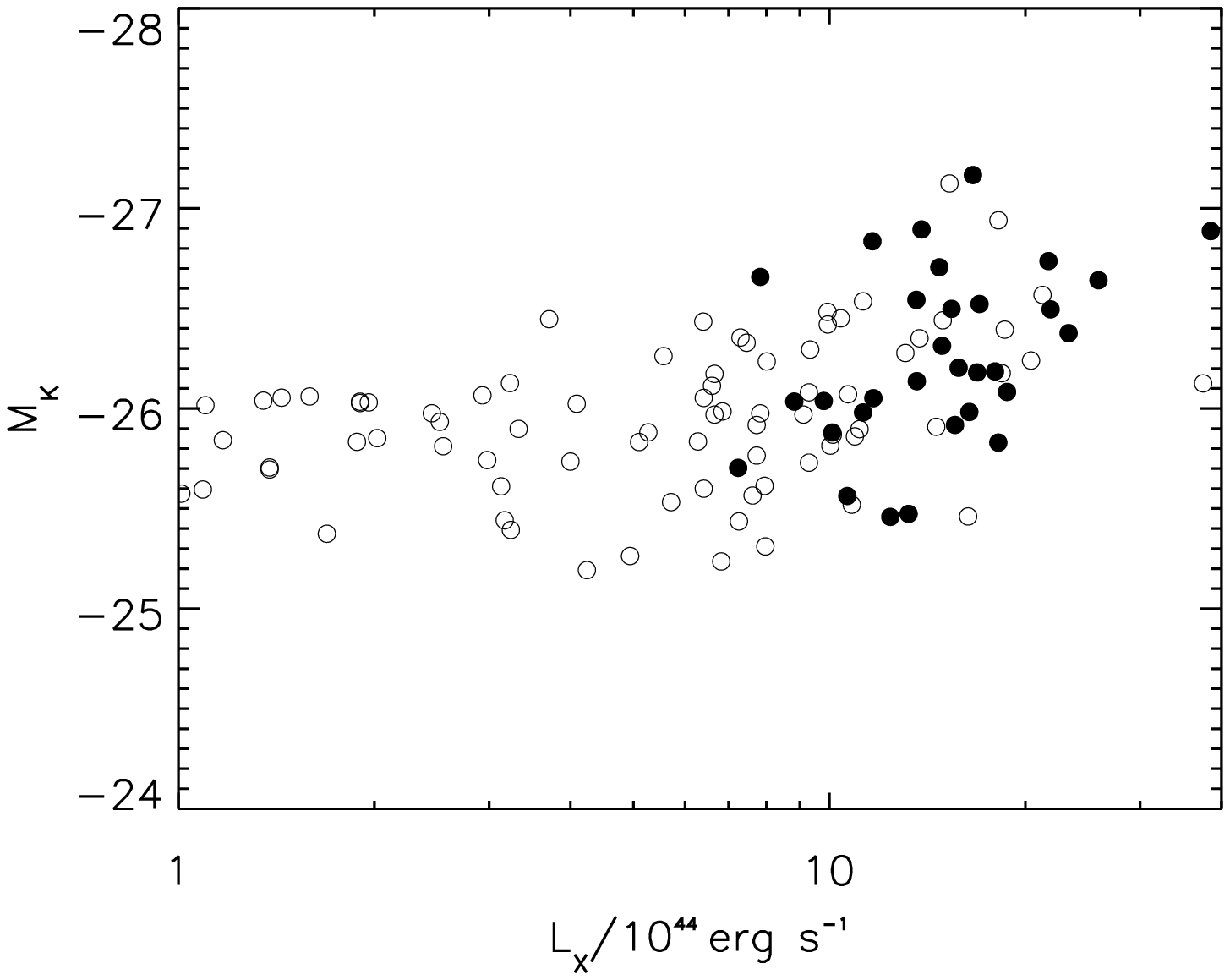}\includegraphics*{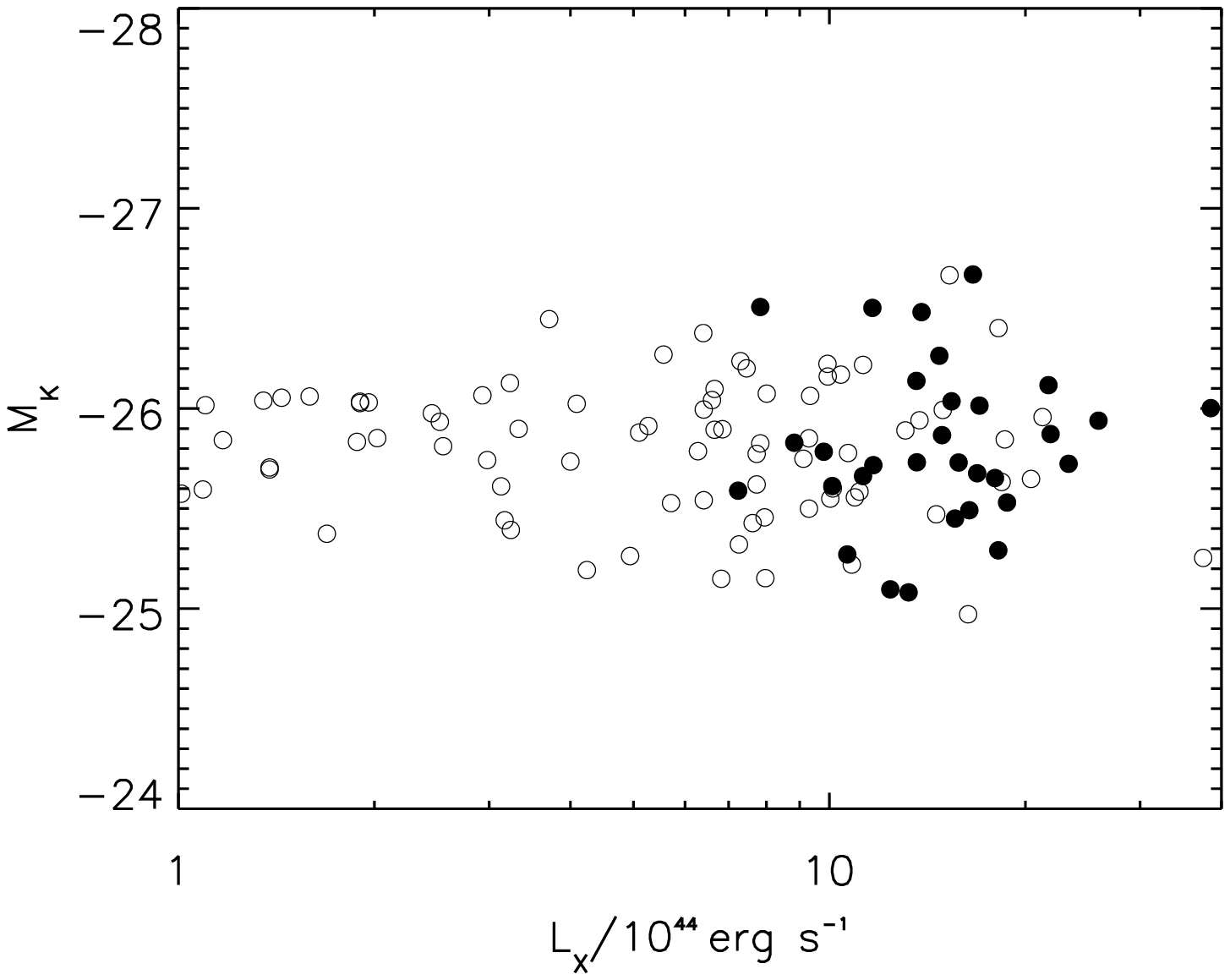}}
\caption[]{{\it Left}: Absolute $K$ band magnitude vs X-ray luminosity.  The unfilled and filled points are BCGs below and above the mean redshift of the sample ($z$=0.25) respectively. {\it Right}: The $L_{X}$ corrected absolute $K$ band magnitude vs X-ray luminosity.}
\label{fig:kx}
\end{figure*}

To observe whether BCGs from the most X-ray luminous clusters can be considered as `standard candles' we present the absolute $K$ band magnitude vs the X-ray luminosity (Fig. \ref{fig:kx} {\it Left}). The absolute $K$ magnitude was calculated using K and passive evolution corrections from a ~\cite{bruzcharl2003} SED with a simple stellar population (SSP), $z_{f}=$5 and solar metallicity. 

From the left panel of Fig. \ref{fig:kx} we find that there is no correlation between absolute magnitude and $L_{X}$ below $L_{X}\sim5\times10^{44}$erg s$^{-1}$, however beyond this value there appears to be a positive relationship. The strength of this correlation is found to be moderate with a Pearson correlation statistic r=0.46. To demonstrate that this is not caused by a difference in the photometric technique between the high and low redshift samples, we highlight the BCGs above and below the mean redshift ($z$=0.25) which shows that the high and low $z$ BCGs in the trend region are well mixed. 

Instead of a $M_{K}$ correction with $L_{X}$ we could assume that the trend in Fig. \ref{fig:kx} is caused by evolution with redshift as our clusters are increasingly more X-ray luminous at higher $z$ (Fig. \ref{fig:xz}). If we do ascribe the trend to redshift then we require $\gtrsim$2 mag of passive evolution to provide the same $M_{K}$ correction as that with $L_{X}$. This magnitude of passive evolution at $z=$1 is not seen in stellar population models so we believe that our $M_{K}$ -- $L_{X}$ trend is real and we concentrate on this for the remainder of the paper.  

Previous works have also found that BCGs from higher $L_{X}$ clusters are brighter in the $K$ band (\citealt{Collins1998}, \citealt{brough2005}).  Observations suggest that BCGs in higher mass systems assemble their stellar mass earlier and are therefore brighter than those from less massive clusters \citep{brough2005}. This is qualitatively consistent with theories of hierarchical assembly.

\begin{figure*}
\centering
\scalebox{0.50}[0.50]{\includegraphics*[bb= 100 230 530 520]{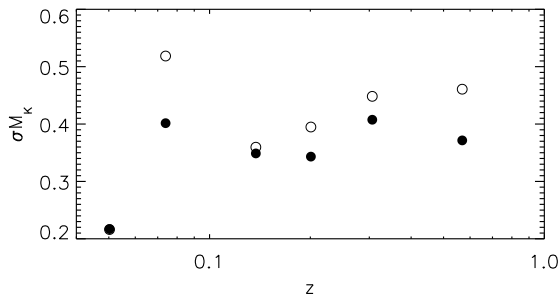}}
\caption[]{The 1$\sigma$ dispersion in Absolute $K$ band magnitude vs redshift. The filled/unfilled points are for the corrected/uncorrected $M_{K}$ values.}
\label{fig:scatterz}
\end{figure*}

We quantify the trend with a two parameter chi square minimised fit to the $L_{X}>5\times10^{44}$erg s$^{-1}$ BCGs. This lower $L_{X}$ limit is chosen as this is the value where the high $L_{X}$ trend with $M_{K}$ intersects with the median $M_{K}$ value of the low $L_{X}$ BCGs. The gradient of this fit is found to be $-1.1\pm0.3$ magnitudes per decade of X-ray luminosity. We then use this fit to correct for the effect of $L_{X}$ on the magnitude, shown in the right panel of  Fig \ref{fig:kx}. In Fig. \ref{fig:scatterz} we plot the 1$\sigma$ dispersion in the absolute magnitude versus $z$ for both the corrected and uncorrected samples. From this we can see that the applied correction reduces the scatter by a factor of $\sim$30\%. 

The mean absolute $K$ band magnitude for the $L_{X}$ corrected cluster sample is $-25.81\pm0.35$ compared to a mean $M_{K}$ of $-26.23\pm$0.45 mag for the uncorrected $K$ band data. The mean magnitude of our sample is therefore comparable to the $-26.40\pm$0.47 mag of~\cite{Collins1998}. For comparison with simulation, \cite{DeLucia2007} find their mean $M_{K}=-26.6\pm$0.16 mag which, although brighter, is within 1 sigma of our uncorrected mean. This demonstrates that the results from our findings can be compared to the work of both  \cite{Collins1998} and \cite{DeLucia2007}.

\subsection{Hubble Diagram}
\label{sec:kz}
\begin{figure*}
\centering
\scalebox{0.50}[0.50]{\includegraphics*{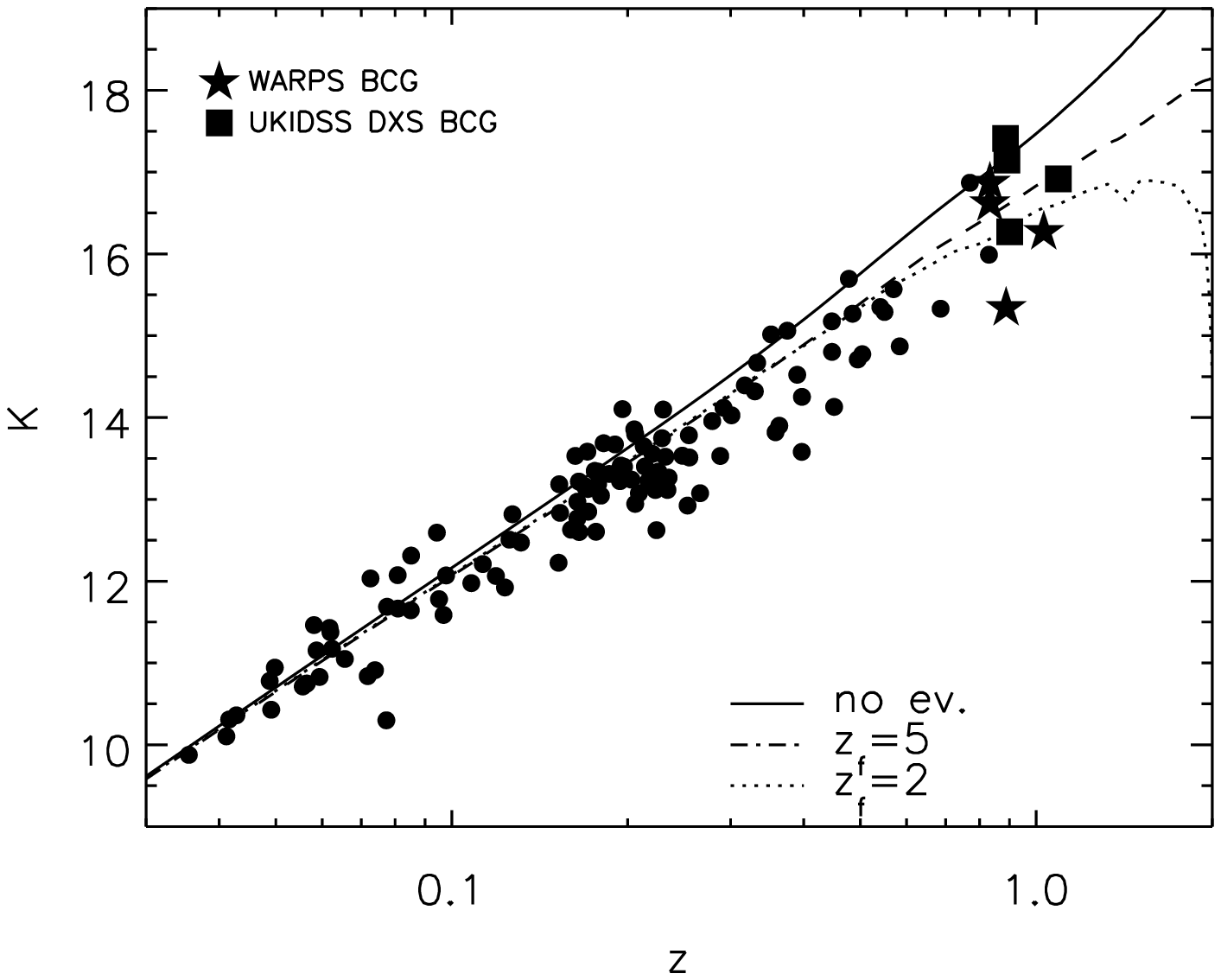}\includegraphics*{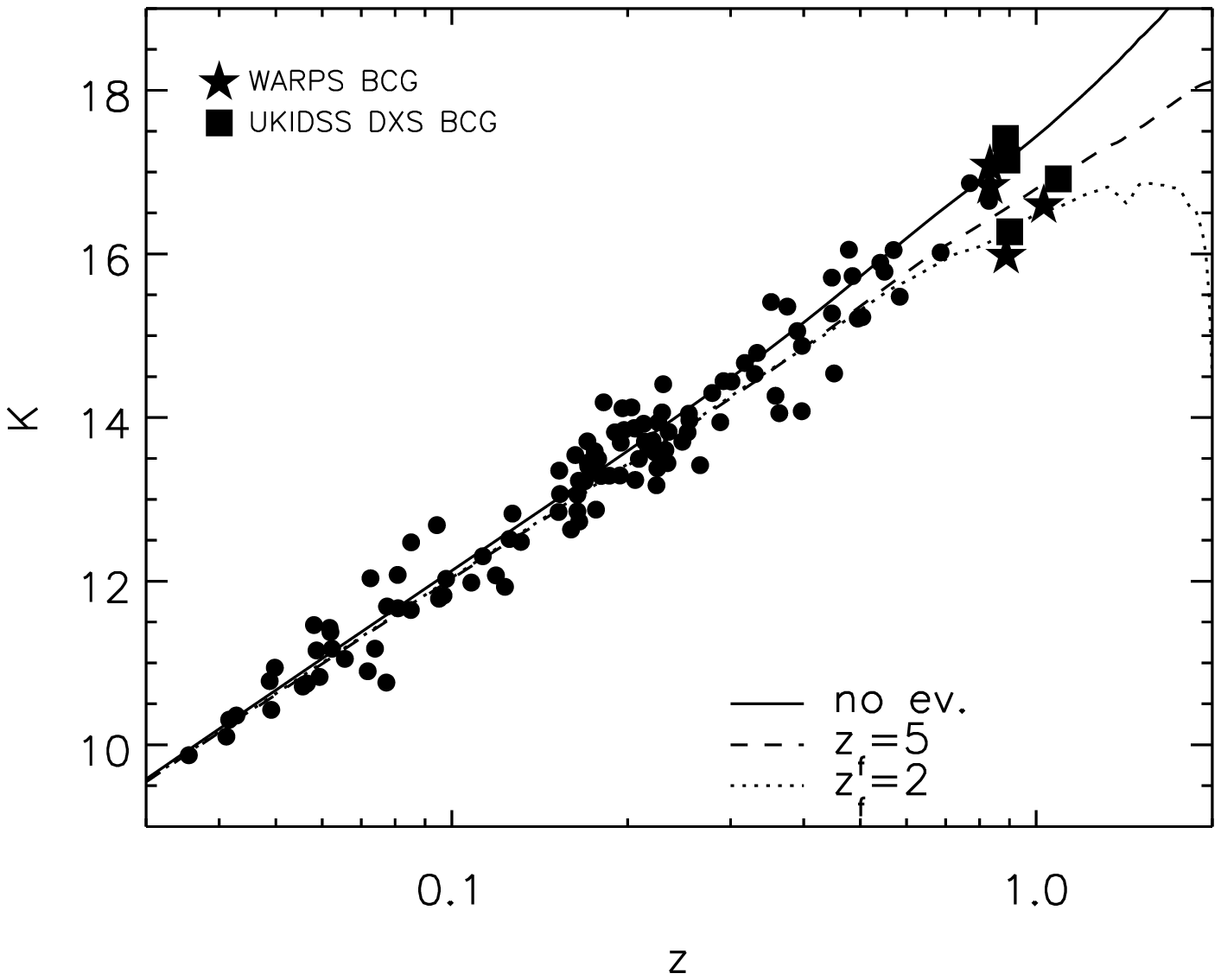}}
\caption[]{{\it Left}: $K$ vs $z$ Hubble diagram for the entire BCG sample. The lines represent different models from the ~\cite{bruzcharl2003} GALAXEV codes. All models assume a Salpeter IMF and Solar metallicity. {\it Right}: The corrected $K$ vs $z$ Hubble diagram for the entire BCG sample.}
\label{fig:kz}
\end{figure*}

Now that we are satisfied that we have limited environmental effects within our sample we can test the nature of the BCG evolution. The Hubble diagram probes both the build up of mass and stellar evolution of the BCGs. Figure \ref{fig:kz} shows the uncorrected and X-ray luminosity corrected $K$ band Hubble diagrams for the whole BCG sample respectively. The uncorrected Hubble diagram is included for comparison to demonstrate the success of the $L_{X}$ -- magnitude correction introduced in \S\ref{sec:xc}. 

The lines plotted represent various stellar population models from the ~\cite{bruzcharl2003} GALAXEV code. All models assume a Salpeter IMF ~\citep{Salpeter1955} and solar metallicity ~\citep{Humphrey2006}. The models are normalised to the median BCG magnitude at $z\lesssim$0.1. The formation redshifts of $z_{f}=$5, $z_{f}=$2 and a no evolution model are chosen for comparison with ~\cite{BCM2000}. 

\begin{figure}
\centering
\scalebox{0.50}[0.50]{\includegraphics*{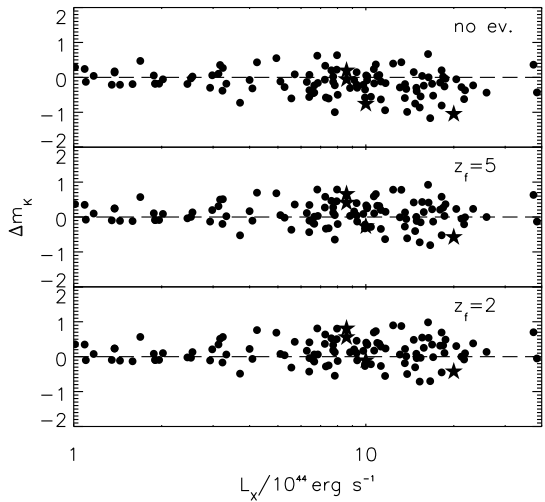}}
\caption[]{The residuals of the BCG corrected $m_{K}$ about the non-evolution and the $z_{f}=$5 and $z_{f}=$2 passive evolution models from Fig. \ref{fig:kz}. $\Delta m_{K}=m_{BCG}-m_{model}$. The stars represent the WARPS BCGs.}
\label{fig:kxres}
\end{figure}

By measuring the residuals about each model track in the corrected Hubble diagram we can identify which scenario best describes the data. Fig. \ref{fig:kxres} shows these residuals. The r.m.s scatters about the three models are 0.44, 0.41 and 0.40 for the no evolution, $z_{f}=$2 and $z_{f}=$5 passive evolution models respectively. We find that the passive evolution models provide a better description than no evolution in agreement with the observations of ~\cite{BCM2000} and the simulations of \cite{DeLucia2007}. 

As the significance of this result is low and there is some uncertainty over the validity of the $M_{K}$ -- $L_{X}$ correction we look to further constrain the evolution of BCGs by investigating $J-K$ colour evolution with redshift in \S\ref{sec:cev}.

\subsection{Colour Evolution with Redshift} 
\label{sec:cev}
\begin{figure}
\centering
\scalebox{0.50}[0.50]{\includegraphics*{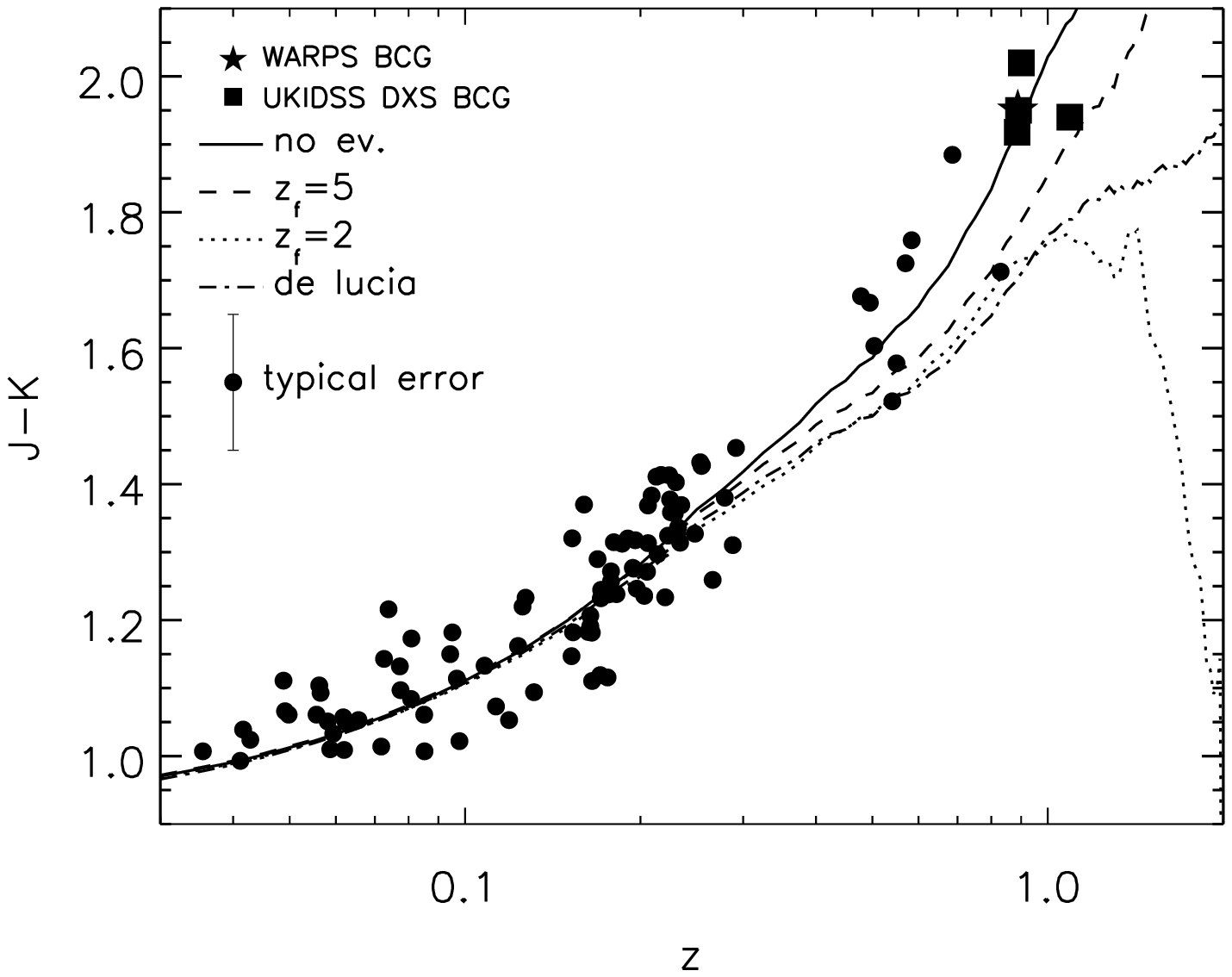}}
\caption[]{J-K vs z for the entire BCG sample. The Lines represent different models from the ~\cite{bruzcharl2003} GALAXEV codes and the hierarchical merger model of \cite{DeLucia2007}. All of the GALAXEV models assume a Salpeter IMF and Solar metallicity. The \cite{DeLucia2007} model forms 50 per cent of the BCG stellar content by $z\sim$5 and 80 per cent by $z\sim$3 with a Chabrier IMF. The WARPS BCG with J band data is from the cluster ClJ1226.9+3332.}
\label{fig:jkz}
\end{figure}

Most BCG studies concentrate on the $K$ band evolution with redshift but here we introduce the $J$ band to observe the evolution of the $J-K$ colour, which probes the stellar evolution of our sample.  Fig. \ref{fig:jkz} is the $J-K$ colour vs $z$ for our BCG sample. As in \S\ref{sec:kz} the models we compare to are calculated using the ~\cite{bruzcharl2003} GALAXEV codes. These models assume a Salpeter IMF and solar metallicity.

In addition, the dash dot line represents the model from \cite{DeLucia2007}. This model forms 50 per cent of the BCG stellar content by $z\sim$5 and 80 per cent by $z\sim$3. We calculate that this corresponds to an exponentially decreasing star formation rate with an $e$-folding time $\tau\sim$0.93Gyr. For consistency we calculate this photometric model using the same population synthesis as \cite{DeLucia2007}, a \cite{bruzcharl2003} model with a \cite{chab2003} IMF.

The data show no real preference for a particular evolution track up to $z\sim$0.4 as the models show little divergence up to this point. However, beyond this redshift the ISAAC, UKIDSS DXS and WARPS BCGs appear to favour the no evolution or passive $z_{f}=$5 models over $z_{f}=$2. We quantify this for the redshift range 0.8$<$z$<$1 by comparing the mean $J-K$ colour to the model values. We find a formation redshift $z_{f}=$2 is ruled out to a significance of 6$\sigma$ while the $z_{f}=$5 and no evolution models are both within 3$\sigma$ of the mean BCG colour. These results are in agreement with both the observations of \cite{BCM2000} who favour $z_{f}>5$ and the no evolution result of \cite{Alfonso1998}. The \cite{DeLucia2007} model is shown to be in good agreement with our observations at low $z$ but becomes too blue compared to our current high $z$ data, suggesting its star formation lasts for too long. We calculate that this model would provide a better description to our data if it had an exponentially decreasing star formation rate with $e$-folding time $\tau\sim$0.5Gyr.

\subsection{Emission Lines}
With the unprecedented wealth of colour information available for our sample in combination with spectroscopy from \cite{Crawford1999} we can investigate the sub-population of BCGs that have line emission. This emission originates from star forming activity attributed to cooling flows (see review by ~\citealt{Fabian1994}). In this process intracluster gas cooling near the cluster centre accretes onto the BCG where it can trigger star formation and therefore line emission. 

BCGs with the most luminous line emission (L(H$\alpha)>10^{41}$erg s$^{-1}$) are found to have a significantly bluer continuum and therefore a bluer optical colour than those with less or no line emission ~\citep{Crawford1999}. Here we look for this same trend in the near-infrared colour which may aid selection of high redshift cooling flow clusters for future studies. 

\begin{figure}
\centering
\scalebox{0.50}[0.50]{\includegraphics*{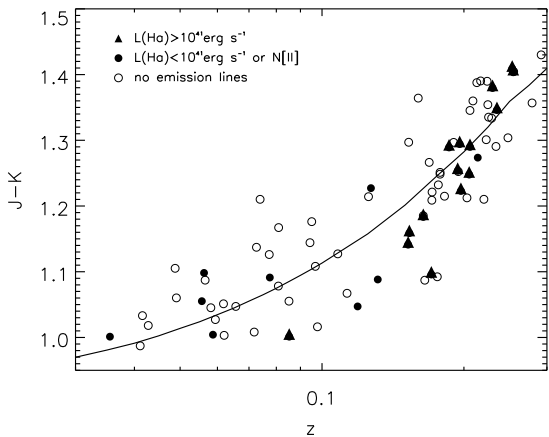}}
\caption[]{$J-K$ vs $z$ for BCGs with spectral information normalised to the solid non-evolution line from figure \ref{fig:jkz}.}
\label{fig:jkzem}
\end{figure}

The ~\cite{Crawford1999} table of the \emph{ROSAT} BCS (~\citealt{Ebeling2000}) contains BCG emission line data for all of the members of the BCS sample. This information is included in our $J-K$ vs $z$ plot to see if there is any trend between line emission and near infrared colour (Fig. \ref{fig:jkzem}). For this plot the sample has been normalised to fit the $J-K$ non-evolution track  from Fig. \ref{fig:jkz} to ensure there are no errors due to an unknown colour term. 

\begin{figure}
\centering
\scalebox{0.50}[0.50]{\includegraphics*{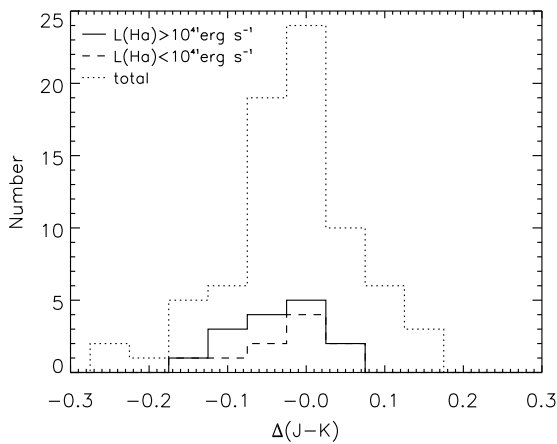}}
\caption[]{This is the histogram of the BCG distribution about the no evolution line. Negative values are blue-ward of the no evolution line, positive values are red-ward.}
\label{fig:spread}
\end{figure}

Fig. \ref{fig:spread} shows a histogram of the BCG distribution about the no evolution line in Fig. \ref{fig:jkzem}. Negative values are blue-ward of the model and positive values are red-ward. The figure shows that we find that both our high (L(H$\alpha)>10^{41}$erg s$^{-1}$) and low luminosity H$\alpha$ emitting BCG populations have a peak in the centre and then a blue tail. This is also seen in optical studies which find a Gaussian around the zero position and a number of populated bins tailing off on the blue side (e.g. ~\citealt{Nat2003}). We can therefore say that we do find a correspondence between the presence of BCG H$\alpha$ emission lines and blue near-infrared colour. In addition we find no correlation between the presence of line emission and the BCG degree of dominance.

\section{Summary}
We have studied the evolution and environment of BCGs in the most X-ray luminous clusters since $z\sim$1.

We find a positive relationship between the near infrared luminosity of the BCG and the X-ray luminosity of its host cluster for clusters where $L_{X}>5\times10^{44} $erg s$^{-1}$. Previous studies have lacked the sample coverage of this work in the high $L_{X}$ regime required to observe this trend. When a correction for this $M_{K}$--$L_{X}$ relation is applied the scatter in the BCG absolute magnitude is reduced.

The $K$ band Hubble diagram for the corrected sample is shown to follow passive evolution. This result is in agreement with the observations of ~\cite{BCM2000} and the simulations of \cite{DeLucia2007}.  

To improve the constraints on BCG evolution we include $J$-band photometry allowing us to compare the $J-K$ colour vs redshift to a set of evolution models. We find that the high redshift BCGs from our MACS, UKIDSS DXS and WARPS data appear to rule out passive evolution with a formation redshift less than 2. We therefore expect that the stellar population of BCGs has been in place since at least redshift 2, in agreement with the observations of \cite{BCM2000}. For comparison with simulation, the \cite{DeLucia2007} model (50 per cent stellar content in place by $z\sim$5 and 80 per cent in place by $z\sim$3) provides a good description to our observations at low $z$ but is too blue compared to our current high $z$ data. This suggests that the simulated BCGs form stars for a longer time period than the observed BCGs. We look to confirm this result in the future with additional high redshift data.

When studying the spectra of individual BCGs we observe a correlation between blue near-infrared colour and the presence of high luminosity (L(H$\alpha)>10^{41}$erg s$^{-1}$) emission lines. Fig. \ref{fig:spread} shows that such emission line BCGs mainly lie on the blue side of the near-infrared colour distribution. This has been seen previously in optical studies (\citealt{Nat2003}) and will be a useful tool in concert with X-ray observations for selecting high redshift cooling flow clusters. 

In conclusion we confirm that near-infrared BCG photometry is a valuable tool for probing the evolution of the bright end of the cluster red sequence. When taken in conjunction with faint end studies (e.g. \citealt{stott2007}a) we can begin to build up a unified picture of cluster evolution where the bright end has been in place since high redshift  while the red sequence is being built up over cosmic time by in falling or transforming galaxies.     

\section{Acknowledgements}

We thank the referee for their useful comments which have improved the clarity of this paper.

We also thank Carlton Baugh, Richard Bower and Sarah Brough for useful discussions. JPS acknowledges support through a Particle Physics and Astronomy Research Council Studentship. GPS acknowledges financial support from the Royal Society.

This publication makes use of data products from the Two Micron All Sky Survey, which is a joint project of the University of Massachusetts and the Infrared Processing and Analysis Centre/California Institute of Technology, funded by the National Aeronautics and Space Administration and the National Science Foundation.

This work is based in part on data obtained as part of the UKIRT Infrared Deep Sky Survey.

The United Kingdom Infrared Telescope is operated by the Joint Astronomy Centre on behalf of the Science and Technology Facilities Council of the U.K.

This work is based partly on observations obtained at the Hale 200 inch telescope at Palomar Observatory.

\bibliographystyle{mn2e}
\bibliography{ref}

\newpage
\appendix
\begin{table*}
\section{The BCG Sample}
\begin{center}
\begin{tabular}{lcccccccc}
\hline
Cluster & ra & dec & $z$ & $J$ & $K$ & $M_{K}$ & $L_X$ & $H_{\alpha}^{\dagger}$\\ 
&\multicolumn{2}{c}{(J2000)}&&&&&($10^{44}$erg\,s$^{-1}$)&\\
\hline
\hline

Abell\,1902 & 14:21:40.53 & +37:17:31.0 & 0.160 & 14.00 & 12.63 & -26.26 & 5.56&n \\ 
Abell\,193 & 01:25:07.62 & +08:41:57.6 & 0.049 & 11.49 & 10.43 & -26.06 & 1.59&n \\ 
Abell\,1930 & 14:32:37.96 & +31:38:48.9 & 0.131 & 13.57 & 12.47 & -26.02 & 4.09&y \\ 
Abell\,1991 & 14:54:31.48 & +18:38:32.5 & 0.059 & 12.16 & 11.15 & -25.71 & 1.38&y \\ 
Abell\,2029 & 15:10:56.13 & +05:44:42.4 & 0.077 & 11.43 & 10.30 & -27.12 & 15.29&n \\ 
Abell\,2034 & 15:10:11.71 & +33:29:11.2 & 0.113 & 13.28 & 12.21 & -25.99 & 6.85&n \\ 
Abell\,2052 & 15:16:44.49 & +07:01:17.7 & 0.035 & 10.88 & 9.88 & -25.93 & 2.52&y \\ 
Abell\,2065 & 15:22:24.02 & +27:42:51.7 & 0.073 & 13.18 & 12.03 & -25.26 & 4.94&n \\ 
Abell\,2072 & 15:25:48.66 & +18:14:09.5 & 0.127 & 14.05 & 12.82 & -25.61 & 3.13&y \\ 
Abell\,2107 & 15:39:39.05 & +21:46:57.9 & 0.041 & 11.10 & 10.10 & -26.02 & 1.10&n \\ 
Abell\,2124 & 15:44:59.03 & +36:06:34.1 & 0.066 & 12.10 & 11.05 & -26.04 & 1.35&n \\ 
Abell\,2175 & 16:20:31.14 & +29:53:27.5 & 0.095 & 12.96 & 11.78 & -26.07 & 2.93&n \\ 
Abell\,2204 & 16:32:46.71 & +05:34:30.9 & 0.152 & 13.37 & 12.23 & -26.57 & 21.25&y \\ 
Abell\,2244 & 17:02:42.50 & +34:03:36.7 & 0.097 & 12.70 & 11.59 & -26.29 & 9.34&n \\ 
Abell\,2254 & 17:17:45.89 & +19:40:48.4 & 0.178 & 14.60 & 13.34 & -25.76 & 7.73&n \\ 
Abell\,2259 & 17:20:09.65 & +27:40:07.9 & 0.164 & 13.96 & 12.77 & -26.17 & 6.66&n \\ 
Abell\,2292 & 17:57:06.69 & +53:51:37.5 & 0.119 & 13.12 & 12.06 & -26.24 & 0.73&y \\ 
Abell\,2345 & 21:27:13.72 & -12:09:46.3 & 0.177 & 13.84 & 12.60 & -26.48 & 9.93&n \\ 
Abell\,2377 & 21:45:57.12 & -10:06:18.7 & 0.081 & 13.16 & 12.07 & -25.44 & 3.17&n \\ 
Abell\,2382 & 21:51:55.63 & -15:42:21.6 & 0.062 & 12.49 & 11.43 & -25.54 & 0.91&n \\ 
Abell\,2384 & 21:52:21.97 & -19:32:48.6 & 0.094 & 13.74 & 12.59 & -25.24 & 6.82&n \\ 
Abell\,2402 & 21:58:28.89 & -09:47:49.7 & 0.081 & 12.84 & 11.67 & -25.85 & 2.02&n \\ 
Abell\,2415 & 22:05:35.49 & -05:32:09.7 & 0.058 & 12.51 & 11.46 & -25.37 & 1.69&n \\ 
Abell\,2426 & 22:14:31.59 & -10:22:26.3 & 0.098 & 13.09 & 12.07 & -25.83 & 5.10&n \\ 
Abell\,2428 & 22:16:15.60 & -09:19:59.7 & 0.085 & 12.71 & 11.64 & -25.98 & 2.45&n \\ 
Abell\,2443 & 22:26:07.93 & +17:21:23.5 & 0.108 & 13.11 & 11.98 & -26.13 & 3.23&n \\ 
Abell\,2457 & 22:35:40.80 & +01:29:05.6 & 0.059 & 11.86 & 10.83 & -26.05 & 1.44&n \\ 
Abell\,2495 & 22:50:19.73 & +10:54:12.8 & 0.078 & 12.78 & 11.69 & -25.74 & 2.98&y \\ 
Abell\,2496 & 22:50:55.85 & -16:24:22.0 & 0.123 & 13.09 & 11.92 & -26.45 & 3.71&n \\ 
Abell\,2589 & 23:23:57.45 & +16:46:38.1 & 0.042 & 11.35 & 10.31 & -25.83 & 1.88&n \\ 
Abell\,2593 & 23:24:20.09 & +14:38:49.7 & 0.043 & 11.38 & 10.36 & -25.84 & 1.17&n \\ 
Abell\,2597 & 23:25:19.72 & -12:07:27.0 & 0.085 & 13.32 & 12.31 & -25.31 & 7.97&y \\ 
Abell\,2622 & 23:35:01.50 & +27:22:20.5 & 0.062 & 12.39 & 11.38 & -25.60 & 1.09&n \\ 
Abell\,2626 & 23:36:30.59 & +21:08:49.8 & 0.057 & 11.84 & 10.75 & -26.03 & 1.96&n \\ 
Abell\,2627 & 23:36:42.10 & +23:55:29.1 & 0.126 & 13.73 & 12.51 & -25.90 & 3.33&n \\ 
Abell\,2665 & 23:50:50.56 & +06:08:58.9 & 0.056 & 11.77 & 10.71 & -26.03 & 1.90&y \\ 
Abell\,2665 & 23:50:50.56 & +06:08:58.9 & 0.056 & 11.85 & 10.74 & -26.03 & 1.90&y \\ 
Abell\,2717 & 00:03:12.98 & -35:56:13.6 & 0.050 & 12.00 & 10.94 & -25.57 & 1.01&n \\ 
Abell\,2734 & 00:11:21.66 & -28:51:15.5 & 0.062 & 12.22 & 11.17 & -25.81 & 2.55&y \\ 
Abell\,376 & 02:46:03.93 & +36:54:18.8 & 0.049 & 11.89 & 10.78 & -25.70 & 1.38&n \\ 
Abell\,399 & 02:57:53.13 & +13:01:51.2 & 0.072 & 11.85 & 10.84 & -26.43 & 6.40&n \\ 
Abell\,401 & 02:58:57.78 & +13:34:57.7 & 0.074 & 12.13 & 10.91 & -26.42 & 9.94&n \\ 
\hline
\end{tabular}
\end{center}
\caption{The $z\lesssim$0.15 2MASS BCGs. $\dagger$ denotes prescence of $H_{\alpha}$ emission. y:
$H_{\alpha}$ emission, n: no $H_{\alpha}$ emission}
\label{tab:sample2m}
\end{table*}

\begin{table*}
\begin{center}
\begin{tabular}{lcccccccc}
\hline
Cluster & ra & dec & $z$ & $J$ & $K$ & $M_{K}$ & $L_X$ & $H_{\alpha}^{\dagger}$\\ 
&\multicolumn{2}{c}{(J2000)}&&&&&($10^{44}$erg\,s$^{-1}$)&\\
\hline
\hline
Abell\,115 & 00:56:00.24 & +26:20:31.7 & 0.197 & 14.65 & 13.40 & -25.91 & 14.59 & y \\ 
Abell\,1201 & 11:12:54.50 & +13:26:08.9 & 0.169 & 14.45 & 13.16 & -25.84 & 6.28 & n \\ 
Abell\,1204 & 11:13:20.52 & +17:35:42.5 & 0.171 & 14.70 & 13.58 & -25.44 & 7.26 & y \\ 
Abell\,1246 & 11:23:58.75 & +21:28:47.3 & 0.190 & 14.99 & 13.67 & -25.57 & 7.62 & n \\ 
Abell\,1423 & 11:57:17.38 & +33:36:40.9 & 0.213 & 15.06 & 13.65 & -25.81 & 10.03 & n \\ 
Abell\,1553 & 12:30:48.94 & +10:32:48.4 & 0.165 & 13.71 & 12.60 & -26.35 & 7.30 & n \\ 
Abell\,1682 & 13:06:45.82 & +46:33:32.9 & 0.234 & 14.43 & 13.12 & -26.54 & 11.26 & n \\ 
Abell\,1704 & 13:14:24.67 & +64:34:32.2 & 0.221 & 14.79 & 13.56 & -25.98 & 7.83 & n \\ 
Abell\,1758 & 13:32:38.59 & +50:33:38.7 & 0.279 & 15.34 & 13.96 & -26.05 & 11.68 & n \\ 
Abell\,1763 & 13:35:20.14 & +41:00:03.8 & 0.223 & 14.44 & 13.11 & -26.44 & 14.93 & n \\ 
Abell\,1835 & 14:01:02.06 & +02:52:43.1 & 0.253 & 14.36 & 12.92 & -26.89 & 38.53 & y \\ 
Abell\,1914 & 14:25:56.64 & +37:48:59.4 & 0.171 & 14.09 & 12.85 & -26.18 & 18.39 & n \\ 
Abell\,1961 & 14:44:31.82 & +31:13:36.7 & 0.232 & 14.86 & 13.52 & -26.11 & 6.60 & ... \\ 
Abell\,2009 & 15:00:19.51 & +21:22:10.6 & 0.153 & 14.02 & 12.83 & -25.97 & 9.12 & y \\ 
Abell\,209 & 01:31:52.51 & -13:36:41.0 & 0.209 & 14.46 & 13.07 & -26.35 & 13.75 & n \\ 
Abell\,2111 & 15:39:41.81 & +34:24:43.3 & 0.229 & 15.10 & 13.75 & -25.86 & 10.94 & n \\ 
Abell\,2163 & 16:15:33.57 & -06:09:16.8 & 0.203 & 14.98 & 13.24 & -26.13 & 37.50 & n \\ 
Abell\,2218 & 16:35:49.39 & +66:12:45.1 & 0.176 & 14.46 & 13.35 & -25.73 & 9.30 & n \\ 
Abell\,2219 & 16:40:19.90 & +46:42:41.4 & 0.226 & 14.70 & 13.34 & -26.24 & 20.40 & n \\ 
Abell\,2254 & 17:17:45.91 & +19:40:49.3 & 0.178 & 14.46 & 13.19 & -25.92 & 7.73 & n \\ 
Abell\,2261 & 17:22:27.24 & +32:07:57.9 & 0.224 & 14.04 & 12.62 & -26.94 & 18.18 & n \\ 
Abell\,2445 & 22:26:55.80 & +25:50:09.4 & 0.165 & 14.40 & 13.22 & -25.73 & 4.00 & ... \\ 
Abell\,2561 & 23:13:57.31 & +14:44:21.9 & 0.163 & 14.71 & 13.53 & -25.39 & 3.24 & ... \\ 
Abell\,291 & 02:01:46.80 & -02:11:56.9 & 0.196 & 15.42 & 14.10 & -25.19 & 4.24 & y \\ 
Abell\,521 & 04:54:06.86 & -10:13:23.0 & 0.248 & 14.86 & 13.53 & -26.24 & 8.01 & n \\ 
Abell\,586 & 07:32:20.26 & +31:38:01.9 & 0.171 & 14.36 & 13.13 & -25.90 & 11.12 & n \\ 
Abell\,661 & 08:00:56.78 & +36:03:23.6 & 0.288 & 14.84 & 13.53 & -26.54 & 13.60 & n \\ 
Abell\,665 & 08:30:57.34 & +65:50:31.4 & 0.182 & 14.92 & 13.69 & -25.46 & 16.33 & n \\ 
Abell\,68 & 00:37:06.82 & +09:09:24.3 & 0.255 & 14.94 & 13.51 & -26.31 & 14.89 & n \\ 
Abell\,750 & 09:09:12.70 & +10:58:27.9 & 0.180 & 14.36 & 13.05 & -26.08 & 9.30 & n \\ 
Abell\,773 & 09:17:53.57 & +51:44:02.5 & 0.217 & 14.64 & 13.22 & -26.28 & 13.08 & n \\ 
Abell\,907 & 09:58:21.98 & -11:03:50.3 & 0.153 & 14.50 & 13.18 & -25.61 & 7.95 & n \\ 
Abell\,963 & 10:17:03.65 & +39:02:52.0 & 0.206 & 14.31 & 12.94 & -26.45 & 10.41 & n \\ 
RX\,J1720.1+2638 & 17:20:10.08 & +26:37:33.5 & 0.164 & 14.18 & 12.97 & -25.97 & 6.66 & y \\ 
RX\,J2129.6+0005 & 21:29:39.91 & +00:05:19.7 & 0.235 & 14.64 & 13.27 & -26.39 & 18.59 & y \\ 
Zw\,1432 & 07:51:25.15 & +17:30:51.8 & 0.186 & 14.62 & 13.31 & -25.88 & 5.27 & y \\ 
Zw\,1693 & 08:25:57.82 & +04:14:48.7 & 0.225 & 14.62 & 13.24 & -26.33 & 7.46 & n \\ 
Zw\,1883 & 08:42:55.99 & +29:27:26.0 & 0.194 & 14.50 & 13.22 & -26.05 & 6.41 & y \\ 
Zw\,2089 & 09:00:36.86 & +20:53:41.2 & 0.230 & 15.50 & 14.10 & -25.52 & 10.82 & y \\ 
Zw\,2379 & 09:27:10.68 & +53:27:33.7 & 0.205 & 15.13 & 13.86 & -25.53 & 5.71 & y \\ 
Zw\,2701 & 09:52:49.22 & +51:53:05.8 & 0.214 & 14.70 & 13.40 & -26.07 & 10.68 & y \\ 
Zw\,348 & 01:06:50.60 & +01:04:10.1 & 0.255 & 15.21 & 13.78 & -26.04 & 9.80 & y \\ 
Zw\,3916 & 11:14:27.43 & 58:22:43.5 & 0.206 & 15.11 & 13.80 & -25.60 & 6.41 & y \\ 
Zw\,5247 & 12:34:17.45 & +09:45:59.4 & 0.195 & 14.69 & 13.41 & -25.87 & 10.12 & n \\ 
Zw\,5768 & 13:11:46.22 & +22:01:37.2 & 0.266 & 14.33 & 13.08 & -26.84 & 11.64 & ... \\ 
Zw\,7215 & 15:01:23.09 & +42:20:39.8 & 0.292 & 15.57 & 14.12 & -25.98 & 11.26 & n \\ 
\hline
\end{tabular}
\end{center}
\caption{The 0.15$\lesssim z\lesssim$0.3 WIRC BCGs. $\dagger$ denotes prescence of $H_{\alpha}$ emission. y:
$H_{\alpha}$ emission, n: no $H_{\alpha}$ emission}
\label{tab:samplew}
\end{table*}

\begin{table*}
\begin{center}
\begin{tabular}{lccccccc}
\hline
Cluster & ra & dec & $z$ & $J$ & $K$ & $M_{K}$ & $L_X$\\ 
&\multicolumn{2}{c}{(J2000)}&&&&&($10^{44}$erg\,s$^{-1}$)\\
\hline
\hline
MACS\,J0018.5+1626 & 00:18:33.68 & +16:26:15.1 & 0.541 & 16.87 & 15.35 & -26.08 & 18.74 \\ 
MACS\,J0025.4-1222 & 00:25:27.44 & -12:22:28.3 & 0.478 & 17.37 & 15.70 & -25.46 & 12.40 \\ 
MACS\,J0150.3-1005 & 01:50:21.24 & -10:05:29.6 & 0.363 & 0.00 & 13.90 & -26.66 & 7.83 \\ 
MACS\,J0257.6-2209 & 02:57:09.78 & -23:26:09.8 & 0.504 & 16.38 & 14.77 & -26.50 & 15.40 \\ 
MACS\,J0329.6-0211 & 03:29:41.68 & -02:11:48.9 & 0.451 & 0.00 & 14.13 & -26.89 & 13.85 \\ 
MACS\,J0404.6+1109 & 04:04:32.71 & +11:08:03.5 & 0.358 & 0.00 & 13.82 & -26.71 & 14.75 \\ 
MACS\,J0429.6-0253 & 04:29:36.14 & -02:53:08.3 & 0.397 & 0.00 & 13.58 & -27.17 & 16.61 \\ 
MACS\,J0454.1-0300 & 04:54:11.13 & -03:00:53.8 & 0.550 & 16.87 & 15.29 & -26.18 & 16.86 \\ 
MACS\,J0647.7+7015 & 06:47:51.45 & +70:15:04.4 & 0.584 & 16.63 & 14.87 & -26.74 & 21.70 \\ 
MACS\,J0744.8+3927 & 07:44:51.98 & 39:27:35.1 & 0.686 & 17.21 & 15.33 & -26.64 & 25.90 \\ 
MACS\,J1359.8+6231 & 13:59:54.32 & +62:30:36.3 & 0.330 & 0.00 & 14.32 & -26.03 & 8.83 \\ 
MACS\,J2050.7+0123 & 20:50:43.12 & +01:23:29.4 & 0.333 & 0.00 & 14.67 & -25.70 & 7.24 \\ 
MACS\,J2129.4-0741 & 21:29:26.35 & -7:41:33.5 & 0.570 & 17.29 & 15.57 & -25.98 & 16.40 \\
MACS\,J2214.9-1359 & 22:14:56.51 & 14:00:17.2 & 0.495 & 16.38 & 14.71 & -26.52 & 17.00 \\ 
MACS\,J2241.8+1732 & 22:41:56.18 & +17:32:12.1 & 0.317 & 0.00 & 14.39 & -25.88 & 10.10 \\ 
MACS\,J2245.0+2637 & 22:45:04.62 & +26:38:05.2 & 0.301 & 0.00 & 14.03 & -26.14 & 13.62 \\ 
RCS0224-0002 & 02:24:00.00 & -0:02:00.0 & 0.770 & 0.00 & 16.87 & -25.35 & 0.70 \\ 
MS1054-0321 & 10:57:00.20 & -03:37:27.4 & 0.830 & 17.70 & 15.99 & -26.38 & 23.30 \\ 
\hline
\end{tabular}
\end{center}
\caption{The $z\gtrsim$0.3 MACS and archival BCGs}
\label{tab:sampleu}
\end{table*}


\end{document}